\documentclass{article}

\usepackage{arxiv}

\usepackage[utf8]{inputenc} 
\usepackage[T1]{fontenc}    
\usepackage{hyperref}       
\usepackage{url}            
\usepackage{booktabs}       
\usepackage{amsfonts}       
\usepackage{nicefrac}       
\usepackage{microtype}      
\usepackage{lipsum}		
\usepackage{graphicx}
\usepackage{natbib}
\usepackage{doi}

\usepackage{amsthm}
\usepackage{txfonts}
\usepackage{algpseudocode}

\usepackage{float}
\usepackage{enumitem}
\usepackage{subcaption}
\usepackage[shortcuts]{extdash}

\title{Multi-task learning for classification, segmentation, reconstruction, and detection on~chest CT scans}

\author{ \href{https://orcid.org/0000-0003-2903-6050}{\includegraphics[scale=0.06]{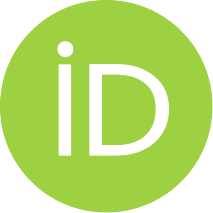}\hspace{1mm}Weronika Hryniewska-Guzik} \\
	Faculty of Mathematics and Information Science\\
	Warsaw University of Technology\\
	Koszykowa 75, 00-662 Warsaw (Poland) \\
	\texttt{weronika.hryniewska.dokt@pw.edu.pl} \\
	\And
    \hspace{1mm}Maria Kędzierska \\
	Faculty of Mathematics and Information Science\\
	Warsaw University of Technology\\
	Koszykowa 75, 00-662 Warsaw (Poland) \\
	\texttt{maria.kaluska.stud@pw.edu.pl} \\
	\And
	\href{https://orcid.org/0000-0001-8423-1823}{\includegraphics[scale=0.06]{orcid.pdf}\hspace{1mm}Przemysław Biecek} \\
	Faculty of Mathematics and Information Science\\
	Warsaw University of Technology\\
	Koszykowa 75, 00-662 Warsaw (Poland) \\
	\texttt{przemyslaw.biecek@pw.edu.pl} \\
}

\renewcommand{\headeright}{} 
\renewcommand{\undertitle}{}
\renewcommand{\shorttitle}{Multi-task learning on chest CT scans}

\hypersetup{
pdftitle={A template for the arxiv style},
pdfsubject={q-bio.NC, q-bio.QM},
pdfauthor={David S.~Hippocampus, Elias D.~Striatum},
pdfkeywords={First keyword, Second keyword, More},
}

\begin{document}
\maketitle

\begin{abstract}
Lung cancer and covid-19 have one of the highest morbidity and mortality rates in the world. For physicians, the identification of lesions is difficult in the early stages of the disease and time-consuming. Therefore, multi-task learning is an approach to extracting important features, such as lesions, from small amounts of medical data because it learns to generalize better. We propose a~novel multi-task framework for classification, segmentation, reconstruction, and detection. To the best of our knowledge, we are the~first ones who added detection to the~multi-task solution. Additionally, we checked the~possibility of using two different backbones and different loss functions in~the~segmentation task.
\end{abstract}



\section{Introduction}
The recent worldwide high contagiousness of the covid-19 virus has stressed the importance of tools that support physicians' work. However, not only covid-19 is the reason why such tools are important. Among cancers, lung cancer has one of the highest morbidity and mortality \cite{cancer}. In the early stages of cancer, due to mild symptoms, it is usually difficult to diagnose \cite{access}. Moreover, physicians are overloaded with work, and the~identification of lesions is very time-consuming.

Computer-aided diagnosis (CAD) systems are designed to assist physicians in interpreting medical images and have to provide the highest possible precision and recall in indicating lesions \cite{8736217}. For this reason, deep learning models seem to be a good solution that meets these requirements. However, the need for responsible solutions that learn correct image features and do not overfit to the~data led to multi-task learning.

Multi-task learning solutions are an approach to extracting important features even from a~small amount of training data, which is common in medical cases. It is a type of learning algorithm that combines information from different tasks (auxiliary tasks) in order to improve the ability to generalize the main task better. In the hard parameter sharing approach, multi-task solutions share some layers and parameters between all the tasks \cite{AMYAR2020104037}.

Various solutions that use multi-tasking for lung medical data have already been developed \cite{AMYAR2020104037,LI2021107848,access}. Amyar et al. \cite{AMYAR2020104037} created a framework based on~the~VGG\=/13 backbone that solved the classification, segmentation, and reconstruction problem. However, after analyzing the publicly available datasets used in that work and the~lack of preprocessing, we can say with a high degree of probability that their model fitted too closely to the selected datasets. 

This paper proposes a novel multi-task framework for classification, segmentation, reconstruction, and detection. We are the first ones who show that it is possible to add detection. Additionally, we checked the possibility of using a~different backbone - ResNet\=/50 and altered the loss function in the~segmentation task. In our solution, we showed that multi-task solution can be extended to new tasks.

\section{Multi-task model training on CT chest scans}

\subsection{Data and preprocessing}

For training and evaluation, the following datasets were used:
\begin{itemize}
\setlength\itemsep{0pt}
    \item 1816 images for classification and reconstruction: non-covid patients from MedSeg \cite{zenodo}, UCSD-AI4H \cite{zhao2020Dataset};
            covid-19 patients from UCSD-AI4H \cite{zhao2020Dataset};
            cancer patients from Lung-PET-CT-Dx \cite{Li2020},
    \item 472 images for segmentation and reconstruction: MedSeg \cite{zenodo} (only images with masks for covid lesions),
    \item 99 images for detection and reconstruction: MedSeg \cite{zenodo} Image masks have 3~possible covid lesions: ground-glass opacity, consolidation, and pleural effusion. 
\end{itemize}

Due to the fact that, in selected CT datasets, not all images were in 3D, we decided to use slices from CT scans, that is, 2D images. In order to unify the~images, we equalized their histograms and rescaled them with their masks to a~size of 256x256. We scaled the pixels to take values in the range $[0, 1]$. Then, images were split into training, validation and testing sets according to Table \ref{splitdata}.

\begin{table}[H]
\caption{Data split into training, validation and testing set.}\label{splitdata}
\centering
\begin{tabular}{|l|r|r|r|}
\hline
Tasks &  Train & Valid & Test\\ \hline
classification \& reconstruction (CR) & 1331 & 244 & 241 \\ 
segmentation \& reconstruction (SR) & 377 & 48 & 47\\
detection \& reconstruction (DR) & 79 & 10 & 10\\
\hline
\end{tabular}
\end{table}

\begin{figure}[h]
    \centering
    \includegraphics[width=\textwidth]{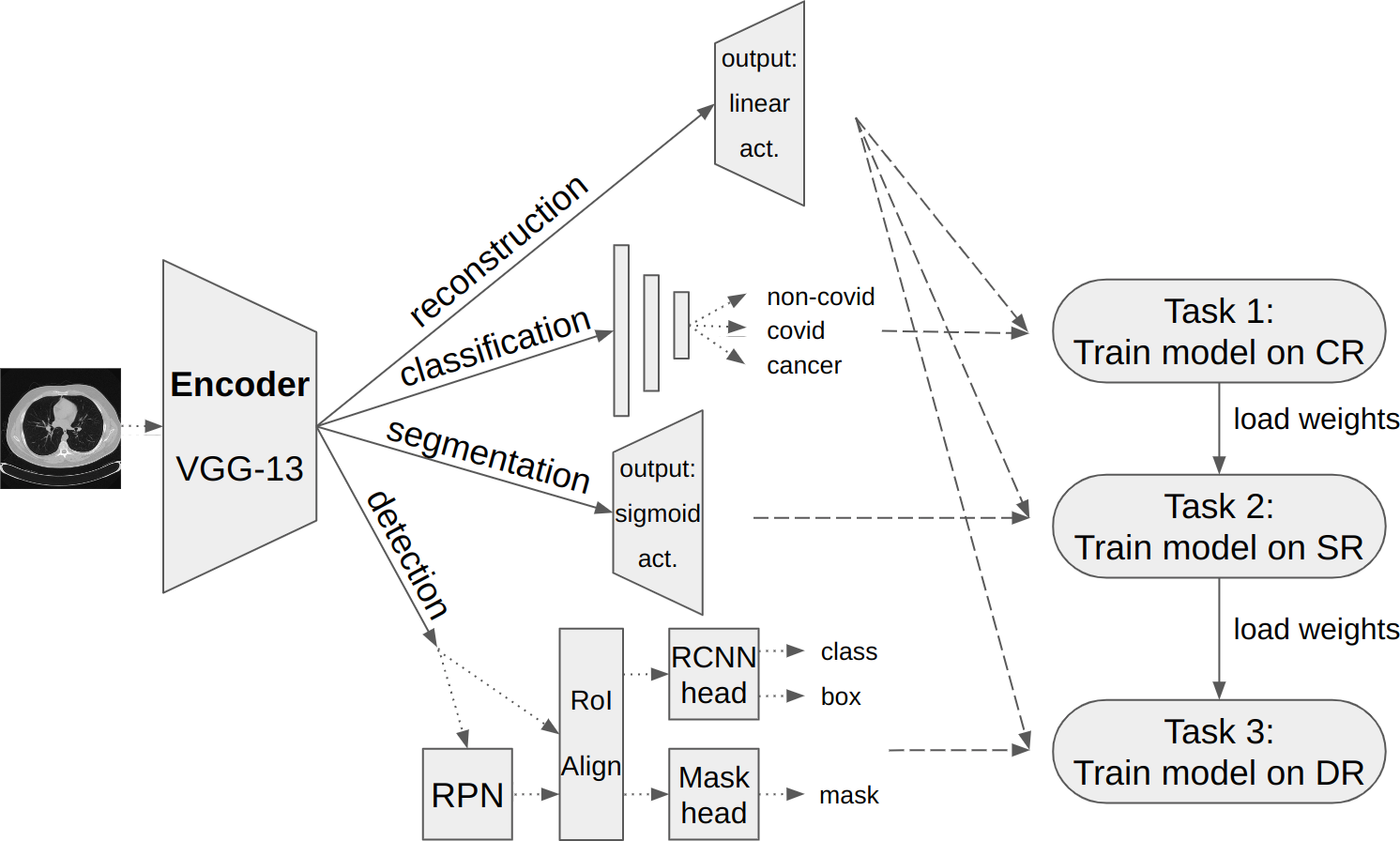}
    \caption{Multi-task architecture and training diagram for tasks: classification (C), segmentation (S), reconstruction (R), and detection (D).}
    \label{fig:arch}
\end{figure}

\subsection{Multi-task architecture}
The proposed multi-task learning architecture is based on 4 tasks: classification, segmentation, reconstruction and detection. As presented in Fig. \ref{fig:arch}, the architecture is based on U-net \cite{unet} architecture, thus, the shared encoder is a VGG\=/13 neural network.

An encoder takes images in size 256x256x1. The reconstruction decoder is the~second half of
U-net has changed the activation function on the~output layer to linear activation. The segmentation decoder is in the form of the standard second half of U-net, which means that it has a sigmoid activation function on the~output layer. The classification decoder consists of 3 fully connected layers with softmax activation on the~output layer. Mask-RCNN \cite{He_2017_ICCV} returns bounding boxes and masks for detected objects.
\begin{equation}
{\displaystyle {\mathcal {L}_{total}}=
w_1\cdot \displaystyle {\mathcal {L}_{classif}}+
w_2\cdot \displaystyle {\mathcal {L}_{segm}}+
w_3\cdot \displaystyle {\mathcal {L}_{recon}}+
w_4\cdot \displaystyle {\mathcal {L}_{detect}}.
}
\label{equation:loss}
\end{equation}

The final loss function is a sum of weighted losses for specific tasks (Equation~\ref{equation:loss}): categorical cross-entropy loss, generalized Dice loss, mean squared error, and mask R-CNN losses. Generalized Dice loss \cite{loss} in the segmentation task takes into account the unbalanced area of the lesion relative to the area of the entire image, therefore, providing better training results. Weights $w_i$ in our case are \{0, 1\}. The mask R-CNN losses is a sum of the following losses:


\begin{equation}
{\displaystyle {\mathcal {L}_{detect}}=
\displaystyle {\mathcal {L}_{MRCNNclassif}}+
\displaystyle {\mathcal {L}_{MRCNNbbox}}+
\displaystyle {\mathcal {L}_{MRCNNmask}}.
}
\label{equation:loss2}
\end{equation}

\subsection{Model training and results}

The training procedure was divided into 3 steps, shown in Fig. \ref{fig:arch}. Firstly, the~model was trained on image reconstruction and multiclass classification tasks.  In order to verify whether the reconstruction task was unnecessary, the model performed only the~classification task. The results of training two tasks simultaneously, presented in~Table \ref{tab:classreconst}, were slightly better than the results of training classification only.

\begin{table}[h]
    \vspace*{-5mm}
    \caption{Performance metrics of the model trained concurrently on two tasks: classification \& reconstruction and on the only classification task.}\label{tab:classreconst}
    \centering
    \setlength{\tabcolsep}{2pt}
    \begin{tabular}{|l|r|r|r|r|r|}
    \hline
    Task & Accuracy &  Macro F1 & F1 non-covid & F1 covid-19 & F1 cancer\\ \hline
    
    Classification \& reconstruction &  0.89 & 0.91 & 0.88 & 0.90 & 0.97 \\ 
    Only classification & 0.87 & 0.89 & 0.83 & 0.89 & 0.97 \\
    \hline
    \end{tabular}
\end{table}

\begin{figure}[h]
    \centering
    \includegraphics[width=\textwidth]{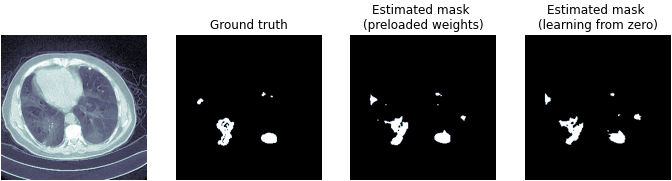}
    \caption{Masks generated by concurrent segmentation \& reconstruction task with and without loading weights from classification task.}
    \label{fig:sr}
\end{figure}

Secondly, the model was given the following tasks: segmentation of covid-19 lesions and image reconstruction.  This was done on a smaller dataset. In~the~first approach, the model had preloaded  weights from the previous task, and in~the~second approach, the network was trained from scratch. The results of~training for 700 epochs with preloaded weights, shown in Table \ref{tab:segmreconst} and Figure \ref{fig:sr}, were better.

\begin{table}[H]
    \caption{Performance metrics of the~model trained on two tasks: segmentation \& reconstruction with and without loading model's weights from the~previous task: classification \& reconstruction. IoU is an abbreviation for Intersection over Union.}\label{tab:segmreconst}
    \centering
    \setlength{\tabcolsep}{2pt}
    \begin{tabular}{|l|r|r|r|r|r|r|r|r|}
    \hline
    Task & Accuracy & F1 & Sensitivity & Specificity & Precision & ROC AUC & IoU\\ \hline
    With loaded weights & 0.99 & 0.78 & 0.76 & 0.99 & 0.80 & 0.88 & 0.64 \\ 
    Without loaded weights & 0.99 & 0.75 & 0.75 & 0.99 & 0.76 & 0.87 & 0.60 \\ \hline
    \end{tabular}
\end{table}

\begin{figure}[h]
    \includegraphics[width=1.\linewidth]{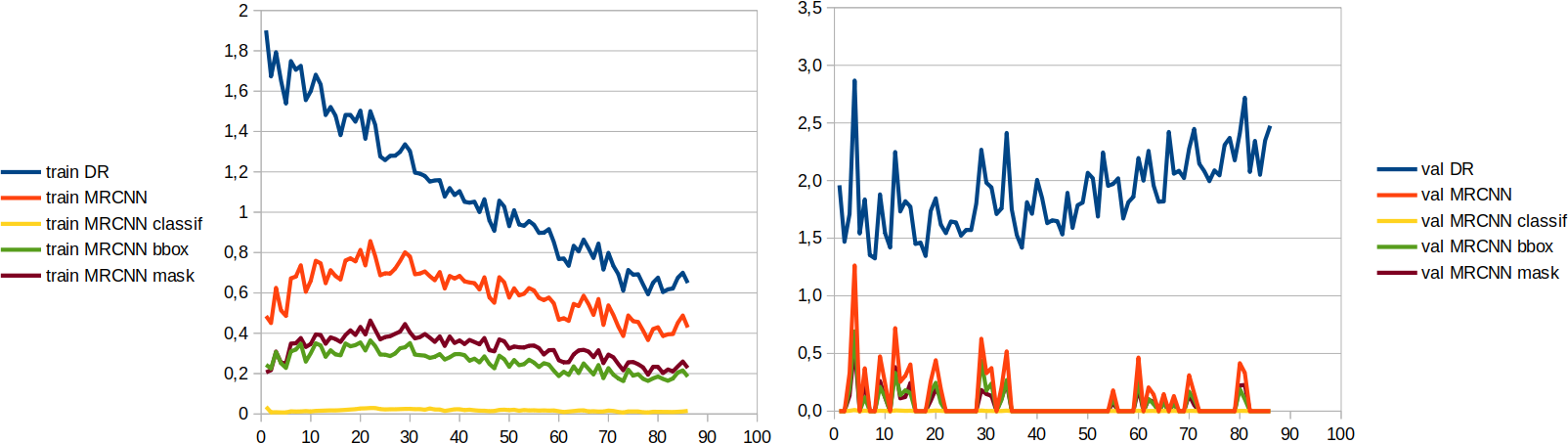}
    \caption{Evaluation of the loss during training detection \& reconstruction task as a function of epoch.}
    \label{DR}
\end{figure}

In order to combine previous networks with mask RCNN, the backbone of MaskRCNN was changed to VGG\=/13. Then, the weights from the best model in the~segmentation \& reconstruction task were loaded. However, due to the small training set, the network was overfitting from the beginning, presented in~Figure~\ref{DR}. To overcome overfitting various data augmentation was applied, such as elastic transformation, rotating by a~small angle, and cropping. Nonetheless, it did not help to obtain satisfactory results.

\begin{figure}[h]
    \begin{subfigure}{1.\textwidth}
      \centering
      \includegraphics[width=1.\linewidth]{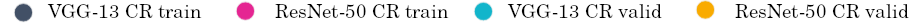}
    \end{subfigure}%
    
    \begin{subfigure}{.33\textwidth}
      \centering
      \includegraphics[width=1.\linewidth]{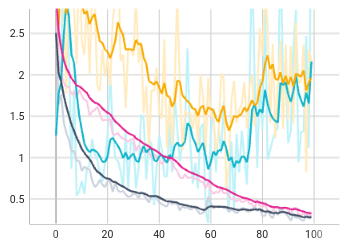}
      \caption{Loss for CR task\\\hspace{\textwidth}}
    \end{subfigure}%
    \begin{subfigure}{.33\textwidth}
      \centering
      \includegraphics[width=1.\linewidth]{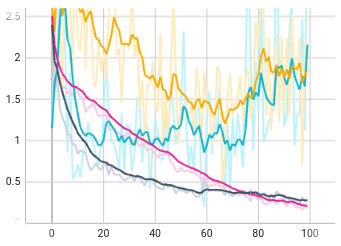}
      \captionsetup{format=hang}
      \caption{Classification loss \\ in~CR~task}
    \end{subfigure}%
    \begin{subfigure}{.33\textwidth}
      \centering
      \includegraphics[width=1.\linewidth]{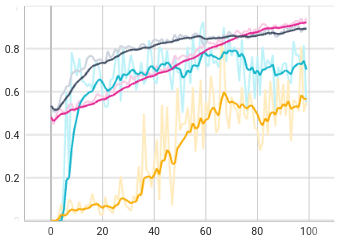}
      \captionsetup{format=hang}
      \caption{Classification accuracy \\ in~CR~task}
    \end{subfigure}
    
    \begin{subfigure}{1.\textwidth}
      \centering
      \includegraphics[width=1.\linewidth]{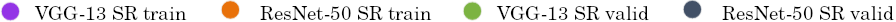}
    \end{subfigure}%
    
    \begin{subfigure}{.33\textwidth}
      \centering
      \includegraphics[width=1.\linewidth]{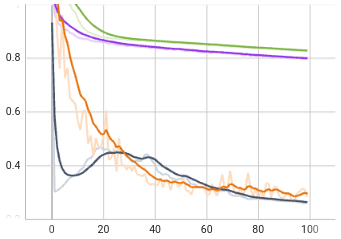}
      \caption{Loss for SR task\\\hspace{\textwidth}}
    \end{subfigure}%
    \begin{subfigure}{.33\textwidth}
      \centering
      \includegraphics[width=1.\linewidth]{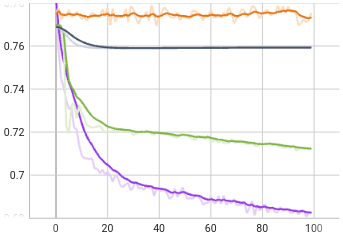}
      \captionsetup{format=hang}
      \caption{Segmentation loss \\ in~SR~task}
    \end{subfigure}%
    \begin{subfigure}{.33\textwidth}
      \centering
      \includegraphics[width=1.\linewidth]{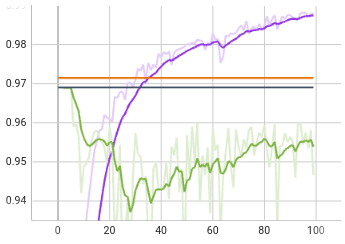}
      \captionsetup{format=hang}
      \caption{Segmentation accuracy \\ in~SR~task}
    \end{subfigure}
    
    \caption{Using two different backbones: VGG\=/13 and~ResNet\=/50 for training classification \& reconstruction (CR) and segmentation \& reconstruction (SR) models. }
    \label{fig:resnet}
\end{figure}

\section{Evaluation on different backbone}

We decided to evaluate whether the~multi-task model obtains similar results on different backbones. Therefore, we change VGG\=/13 backbone to ResNet\=/50, which is the~default backbone in the detection task. We trained a model for 100 epochs on~two different tasks: classification \& reconstruction and segmentation \& reconstruction. The results in Fig.~\ref{fig:resnet} show that there is no strong advantage of one backbone over another. Multi-task model loss is lower in classification \& reconstruction when the~backbone is VGG\=/13, while in segmentation \& reconstruction, the multi-task model loss is lower for backbone ResNet\=/50.

\section{Conclusions}

The framework was successfully created and tested. Obtained classification and segmentation results are satisfactory, especially due to the fact that segmentation applies to small lesions, not whole lungs. However, much more data is needed to get desired results in the~detection task, even in a multi-task approach.


\section*{Acknowledgment}

This work was financially supported by the Polish National Center for Research and Development grant number INFOSTRATEG-I/0022/2021-00, and carried out with the support of the Laboratory of Bioinformatics and Computational Genomics and the High Performance Computing Center of the Faculty of Mathematics and Information Science, Warsaw University of Technology.

\small
\bibliographystyle{unsrtnat}
\bibliography{references}

\end{document}